\documentstyle[prl,aps,psfig]{revtex}
\begin{document}
\begin{center}
{\Large\bf The random field critical concentration in dilute antiferromagnets}

\vspace{0.25in}

{\large W. C. Barber and D. P. Belanger}\\
{Department of Physics, University of California,
Santa Cruz, CA 95064}\\
\end{center}
\begin{abstract}
Monte Carlo techniques are used to investigate the equilibrium
threshold concentration, $x_e$, in the dilute anisotropic
antiferromagnet $Fe_xZn_{1-x}F_2$ in an applied magnetic field,
considered to be an ideal random-field Ising model system.
Above $x_e$ equilibrium behavior
is observed whereas below $x_e$ metastability and domain
formation dominate.  Monte Carlo results agree
very well with experimental data obtained using this system.

\end{abstract}
\vspace{0.4in}

The dilute antiferromagnet (AF) $Fe_xZn_{1-x}F_2$ in an applied field
is a realization\cite{fa79,c84} of the random-field
Ising model (RFIM).
Many studies\cite{b98} have been done on this system for $x<0.75$.  In such
cases, metastability and domain formation mask the equilibrium
critical behavior, particularly in scattering measurements that are by
nature dominated
by long-range antiferromagnetic correlations.  The specific heat, $C_m$,
on the other hand, is not as greatly affected by domain formation except
very close to the transition, $T_c(H)$,
since it is primarily sensitive to short range correlations\cite{sb98}.  Only
recently has it been discovered\cite{sbf99} that equilibrium scattering behavior
can be observed for $x=0.93$ with no evidence of domain formation. 
Domain walls form with little energy cost when vacancies are
so numerous that magnetic bonds can be largely avoided.  At
high magnetic concentrations domain walls must cut a large number
of magnetic bonds and long-range order (LRO) is stable, as in the
Imry-Ma domain wall energy arguments\cite{im75}.  Hence, the RFIM
can be studied in equilibrium for $x=0.93$ and a transition to
LRO is observed, consistent with theory\cite{bk87}. 
The question remains as
to the nature of the disappearance of the domain walls as
the magnetic concentration increases.  Do they disappear gradually
or is there a critical concentration above which they do not form?

We have performed Monte Carlo (MC)
simulations of the RFIM modeled as closely as possible after
$Fe_xZn_{1-x}F_2$.  We provide evidence that
there is a critical equilibrium threshold concentration, $x_e$, above which
domain formation does not occur.  We estimate this concentration to be
$0.75<x_e<0.80$.
The simulation results also shed light on the
experimentally observed\cite{bwshnlrl96} formation of domain walls upon approach
to $T_c(H)$ for $x<x_e$ after cooling in zero field and
subsequently heating in a
field (ZFC).  Finally we demonstrate that cooling in a field (FC) for
$x<x_e$ results in a metastable state at low $T$.

The staggered magnetization ($M_s$) and correlation functions for the first
three nearest-neighbor (NN) pairs have been calculated using
MC simulations. The magnetic lattice
corresponding to the body-centered-tetragonal $Fe_{1-x}Zn_{x}F_2$ lattice
is described as two cubic
sub-lattices of size $L \times L \times L$ each, delineated as one dimensional
arrays bit coded to accommodate large lattice sizes.
Most of the results reported here were obtained with
$L=64$, corresponding to more than $5.2 \times 10^{5}$ sites
magnetically occupied with probability $x$.  Periodic boundary
conditions are used.
At each temperature magnetic sites are visited randomly with each site
visited an average of once per MC step.  Upon each visit, a spin
is flipped with a probability given by the metropolis
algorithm.  The temperature was changed in steps of
$0.01$~K after
$N$ MC steps.  $N$ was increased until the results of the simulation
were largely independent of $N$ for $H=0$.  Unless otherwise stated, we used
$N=500$ for the results reported here.  Using $N=1000$ gave essentially
identical results for $H=0$.
The temperature scanning procedures correspond to the experimental ones,
i.e.\ cooling and heating for $H=0$, and FC and ZFC for $H=13$~T.
$Fe_{1-x}Zn_{x}F_2$ is well represented\cite{hrg70} by the
Hamiltonian
\begin{equation}
H=\sum _{<ij>}J_{ij}\vec{S}_i \cdot \vec{S}_j -D\sum _i {\vec{S}_i}^2
-\vec{H} \cdot \sum _i \vec{S}_i \quad . 
\end{equation}
In the simulations we use the Hamiltonian
\begin{equation}
H=\sum _{<ij>}J_{ij}{S_i}^z {S_j}^z  -H\sum _i {S_i}^z \quad ,
\end{equation}
which corresponds to the Ising limit $D \rightarrow \infty$.
The first three NN exchange interaction strengths are taken from
spin-wave dispersion
measurements\cite{hrg70} on $FeF_{2}$: $J_{1}=-0.069$~K;
$J_{2}=5.278$~K; and $J_{3}=0.279$~ K.
All other interactions are negligible and are not included
in the simulations.  The transition temperature increases
with anisotropy in this system\cite{bkj82}.  Hence, the
infinite anisotropy of the model results in a transition
temperature much higher than in the real system.
As is done elsewhere\cite{rc98}, the included interactions were all
scaled by a factor $2/3$ to
make the transition temperatures more closely correspond to
the ones observed\cite{fkj91} in $Fe_{1-x}Zn_{x}F_2$.

Figure 1 and its inset show typical behavior for $x = 0.60 < x_e$ for
three cases: $H=0$, which is free of hysteresis; FC with $H=13$~T; and ZFC with
$H=13$~T.  In the inset, the behavior of $C_m=dE/dT$ vs.\ $T$ is shown.
For $H=0$ random-exchange behavior is observed\cite{b98} with slight rounding
from finite sample size effects.  Taking the rounding into
consideration\cite{bkfj87}, we estimate the transition temperature
$T_N \approx 45.0$~K.  For the ZFC and FC procedures,
the specific heat is depressed in
$T$ as expected.  Away from the shifted transition,
the two behaviors are essentially identical with a shape more
symmetric than for $H=0$, as is observed in experiments\cite{sb98}.
Since the ZFC and FC behaviors
are the same outside the distorted regions, they clearly share the
same $T_c(H)$.  On a plot of the specific heat versus the logarithm
of the reduced temperature, $T_c(H)$ appears close to the peak
of the ZFC data, $\approx 40.5$~K.   From the main
figure, we see that
for $H=0$, $M_s$ falls to zero very near the $T_N$ taken from $C_m$. 
Upon FC, however, $M_s$ remains unusually small quite far below the
$T_c(H)$ determined from $C_m$.  Upon further cooling, the system
begins to rapidly evolve toward AF LRO with a curvature opposite
to that for $H=0$.  Finally, at a rather well defined
temperature ($T_1$) the $M_s$ curve abruptly changes
curvature.  Upon further cooling, $M_s$ increases
smoothly and slowly, much like the $H=0$ case but shifted to lower $T$.
Upon ZFC, $M_s$ vs.\ T follows a much smoother evolution but nevertheless
approaches $T_c(H)$ with a curvature opposite to the $H=0$ case.
The unusually rapid disappearance
of AF LRO well below $T_c(H)$ is attributed to the formation of domains and
has been observed experimentally\cite{bwshnlrl96}.
The FC behavior in experimental systems, on the other hand,
is quite different from that shown in Fig.\ 1.  In the experiments,
the domain formation prevents significant $M_s$
down to low $T$.  Hence, we examined this behavior further using the MC
techniques.  We found that by slowing the temperature scans by using
$N=1000$ with $L=64$, the domains that form for $x<x_e$
remain stable to much lower
temperatures.  We also did a simulation with $L=128$ and $N=500$ and
again the domains were stable to lower temperatures.  Extrapolating these
two trends to the much larger experimental systems that are cooled much
more slowly relative to the spin flipping time,
it is not surprising that the experimental systems show metastable
domains that are very stable at low temperatures.

Figure 2 shows the ZFC and FC behavior for $x=0.60$, $0.70$, $0.80$ and
$0.90$ at $H=13$~T.  It is clear that the hysteresis
decreases as $x$ increases and is absent
for the cases $x \ge 0.80$.  It thus appears that the systems at larger
$x$ are in equilibrium.  To demonstrate that
there is a critical threshold concentration above which equilibrium behavior,
such as that observed for $x \ge 0.80$, prevails, we plot in Fig.\ 3
the temperature $\Delta T =T_2(H,x) - T_1(H,x)$, where $T_2(H,x)$
is the temperature at which the ZFC joins the FC one just below
$T_c(H)$.  Note that the ZFC and FC curves separate at
$T_1(H)<T_2(H)$.  It is clear that
$\Delta T$ becomes zero somewhat above $x=0.70$ and below $x=0.80$,
strongly indicating a critical concentration $0.75<x_e<0.80$, which
is much larger than the magnetic percolation threshold $x_p=0.25$.
The field dependence of $x_e$ is weak.  Within the accuracy of the
present MC results, we find that for $H=7$~T and $H=13$~T,
$0.75<x_e<0.8$.  This suggests that the critical concentration is
primarily geometric in origin.
Domain formation in the ZFC procedure indicates that for $H>0$, the
free energy must favor domain formation close to $T_c(H)$ for $x<x_e$.
Such behavior is indicated in local-mean-field
simulations\cite{hb84}.  However, there is no evidence for a latent
heat associated with the temperature at which the free energy difference for
the domain state and AF state switches sign.  

Finally, we demonstrate in the inset of Fig.\ 3 that the ZFC peak in
$d({M_s}^2)/dT$ (the Bragg scattering) vs.\ $T$ for $x=0.60$ is at a lower
temperature by more than $1$~K than $T_c(H)$ as determined from
the $C_m$.  This behavior is in agreement
with neutron scattering results\cite{bwshnlrl96} but in
contradiction to the so-called `trompe l'oeil' phenomenology\cite{bfhhrt95}.

In conclusion, we have presented MC evidence for the existence of a
critical concentration $x_e$ above which equilibrium RFIM behavior
prevails and below which domains form close to the transition.
The numerous vacancies for $x<x_e$ allow domain formation without
a great energy cost.  We note that the vacancies themselves percolate
at $x=0.75$ and surely $x_e$ is related to this percolation threshold, although
this may not be the entire story.  Domain walls take advantage of surfaces
of vacancies, not just the filiamentary percolation threshold structures.
Hence, it would not be surprising if $x_e$ were somewhat different from the
vacancy percolation threshold.  Further simulations are currently
under way at $L=128$ and should help to determine $x_e$
more precisely as well as determine the influence of the finite
sample size on the present results.  Certainly, critical behavior
studies should be performed in the real systems only for $x>x_e$.

We acknowledge David Stafford for discussions concerning fast
algorithms for large lattice sizes.
This work has been supported by DOE Grant No. DE-FG03-87ER45324.

\newpage

\begin{figure}[t]
\centerline{\hbox{
\psfig{figure=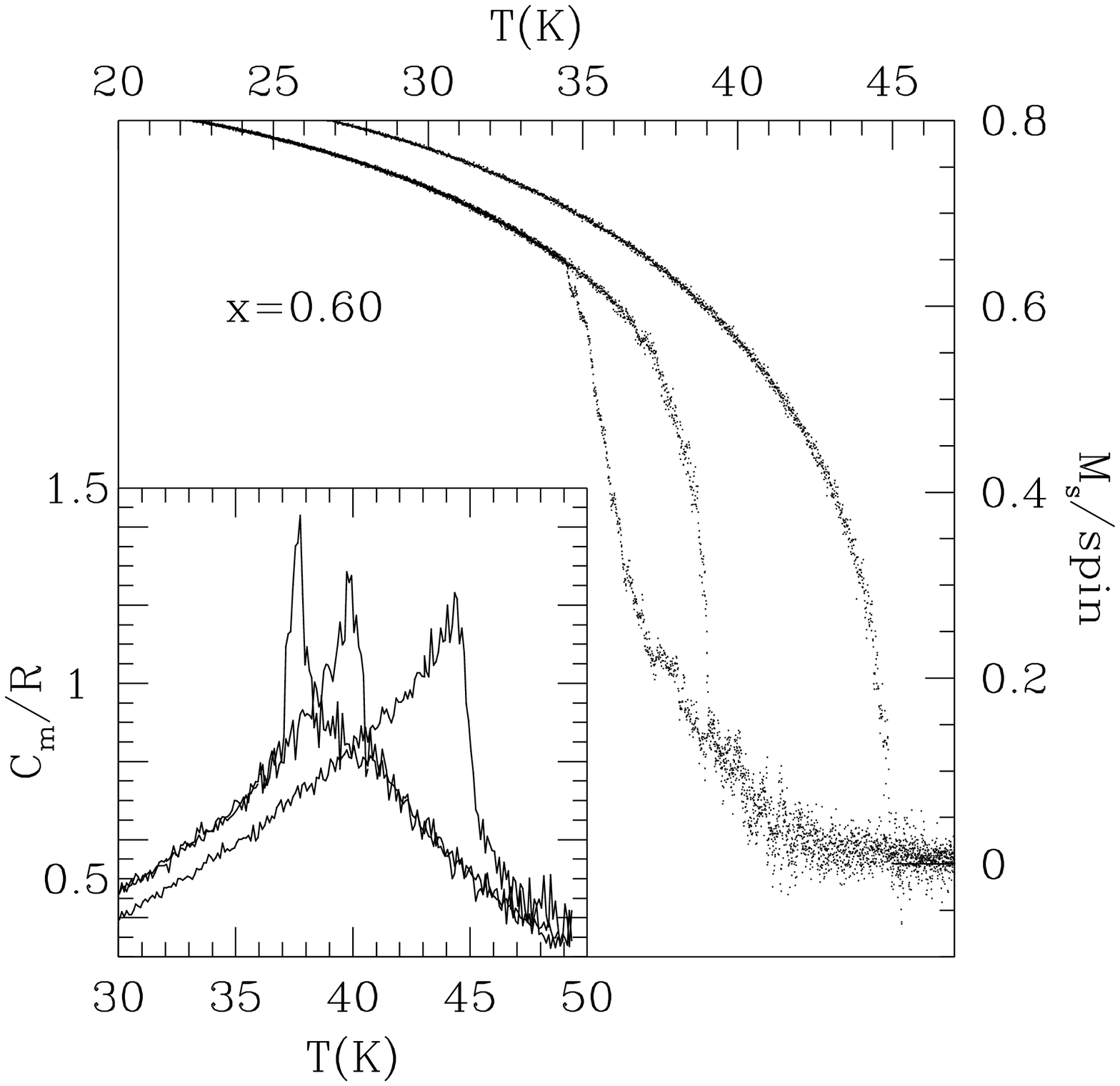,height=5.0in}
}}
\caption{$M_s$ vs.\ $T$ for $x=0.60$ at $H=0$, and
for ZFC and FC at $H=13$~T.  The transition temperatures are determined
from the specific heat behaviors shown in the inset where the units
are expressed in terms of two magnetic sites per unit cell.  In both the
main figure and inset, curves are, from left to right, for the
FC, ZFC and $H=0$ procedures.
}
\end{figure}

\begin{figure}[t]
\centerline{\hbox{
\psfig{figure=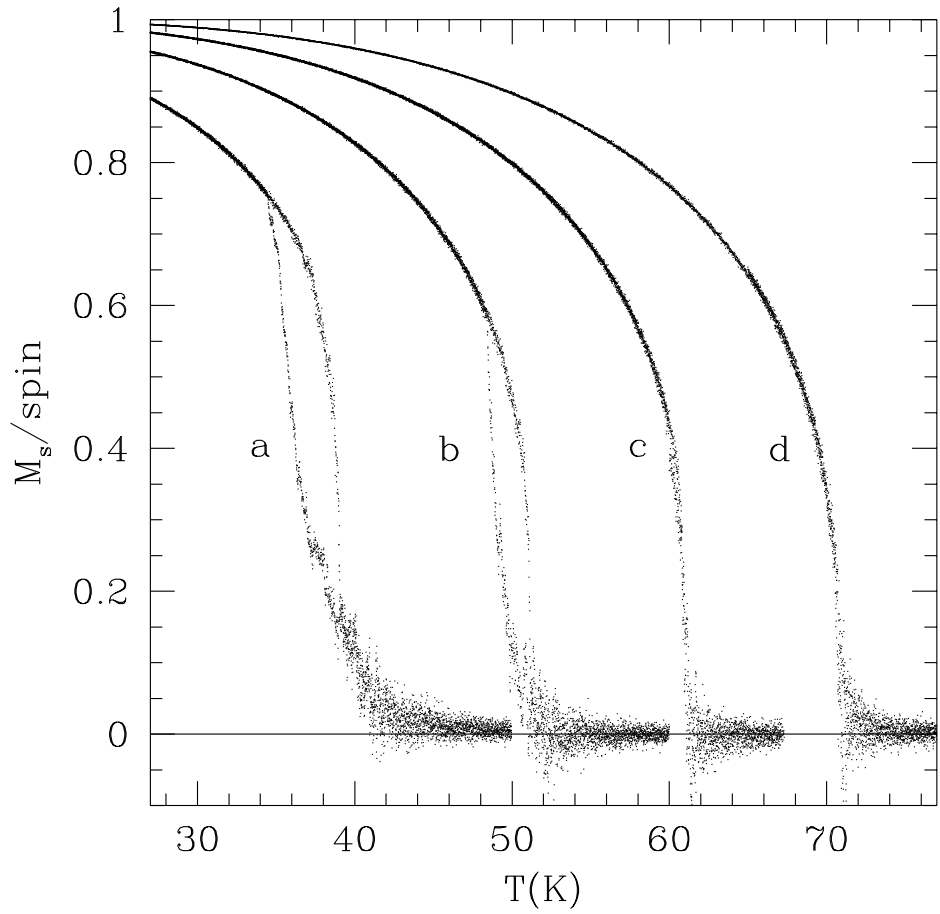,height=5.0in}
}}
\caption{The ZFC and FC $M_s$ vs.\ $T$ for $x=0.60$ (a),
$0.70$ (b), $0.80$ (c) and $0.90$ (d).
 The upper curves are ZFC in each case except $x=0.80$ and $0.90$,
where the ZFC and FC curves coincide at all temperatures.
}
\end{figure}

\begin{figure}[t]
\centerline{\hbox{
\psfig{figure=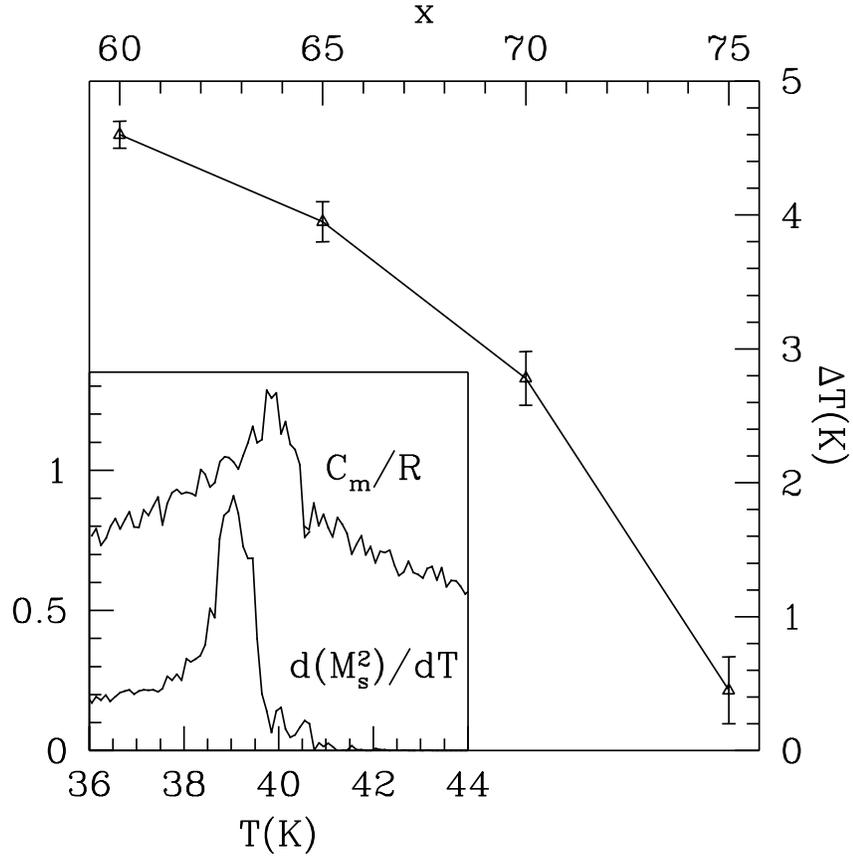,height=5.0in}
}}
\caption{The temperature range over which hysteresis occurs at $H=13$~T,
$\Delta T =T_2(H,x)-T_1(H,x)$ vs.\ x.
The inset shows $d({M_s}^2)/dT$ (arbitrary units) and $C_m/R$
vs.\ $T$ for $x=0.60$.
Note that the peak in $d({M_s}^2)/dT$ vs.\ $T$ is significantly lower than
the corresponding peak in $C_m/R$.
}
\end{figure}

\end{document}